\documentstyle[aps,epsfig,preprint]{revtex}
\tightenlines
\begin{document}
{\title{Coherent coupling of two quantum dots embedded in an Aharonov-Bohm ring}
\draft
\author{A.W. Holleitner$^*$, C.R. Decker, K.~Eberl$^{\dagger}$, and R.H. Blick} 
\address{Center for NanoScience and Sektion Physik,
Ludwig-Maximilians-Universit\"at, Geschwister-Scholl-Platz 1, \\
80539 M\"unchen, Germany.
\\$^{\dagger}$ Max-Planck-Institut f\"ur Festk\"orperforschung,
Heisenbergstr. 1, 70569 Stuttgart, Germany.} 

\date{\today} 
\maketitle
\begin{abstract}
We define two laterally gated small quantum dots ($\sim 15$ electrons) 
in an Aharonov-Bohm geometry
in which the coupling between the two dots can be broadly changed.
For weakly coupled quantum dots we find Aharonov-Bohm oscillations.
In an intermediate coupling regime we concentrate on the molecular states of the double dot
and extract the magnetic field dependence of the coherent coupling.
\pacs{85.30.Vw, 03.65.Bz, 73.40.Gk}
\end{abstract} 
\narrowtext
\newpage 
Quantum dots are the perfect experimental tool for investigating phase
coherence processes in mesoscopic devices~\cite{dot,coherence}.
One of the questions which can be considered is the
entanglement of fermionic particles, e.g. electrons in a solid state environment. 
In this work we present an experimental approach allowing to coherently couple
two quantum dots by a tunneling barrier embedded in an Aharonov-Bohm (AB) ring~\cite{AB}.
For such a system it is expected that singlet and triplet states 
have distinct AB-phases ~\cite{Loss00}. 
Therefore, this setup is a promising candidate for realizing a 
quantum bit in a solid state device~\cite{qubits}.
The fundamental question being addressed is whether 
the coherent coupling of two quantum dots in the few electron limit
can be understood in terms of only two excess electrons, one in each quantum dot, 
or whether the whole shell structure has to be taken into account.
First, we characterize the double quantum dot and 
demonstrate that the coupling between the two quantum dots, 
containing about 15 electrons each, can be varied in a wide range. 
For the case of weak coupling we detect AB-oscillations. 
Secondly, we concentrate on the molecular states of the double quantum dot, 
first evidence of which has been found in transport~\cite{coupling} 
and microwave spectroscopy~\cite{cohmode}.
We extract the magnetic field dependence of the coherent coupling of the
two quantum dots and compare it to recent theoretical models~\cite{Burkhard,QDM}.

The device we used is realized within a two-dimensional electron
gas (2DEG) being $90$~nm below the surface
of an AlGaAs/GaAs heterostructure (see Fig.~1).
At a bath temperature of 35~mK the electron mobility and density are
found to be $\mu = 80$~m$^{2}$/Vs and $n_{s} = 1.7 \times
10^{15}$~m$^{-2}$, respectively. 
By electron beam writing and Au-evaporation Schottky-gates are
defined. Under appropriate voltage bias these form two quantum dots.
As seen in the scanning electron microscope
micrograph of Fig.~1(a) the two gates $gate_1$ and $gate_2$,
which define the center tunneling barrier between the two quantum dots, are
patterned on an additional resist layer. This $45$~nm thick layer is
fabricated from negative resist {\it Calixarene} (hexaacetate {\it
p}-methycalix[6]arene)~\cite{Fujita96}.
Since the dielectric constant of {\it Calixarene} is about
$\epsilon_{Cax} \sim~7.1$~\cite{Vogel} there is no depletion below $gate_1$ and 
$gate_2$~\cite{QPC,remark1}. By aligning these 'carbon-bridges'
to a precision of $\delta x  \le~15$~nm  we obtain an experimental setup
in which one electron can either tunnel
through $dot_1$ or $dot_2$ (see Fig.~1(b)).

By setting the lead-dot coupling of one of the quantum dots to be 
$\Gamma_{lj} = \Gamma_{rj} \cong 0~\mu$eV ($j = 1, 2$ as in Fig.~1(b)), 
we first characterize each dot individually.
From transport spectroscopy we find the following charging energies for each 
dot $E^{dot1}_C = e^2/2C^{dot1}_{\Sigma} = 1.68$~meV and
$E^{dot2}_C~=~1.71$~meV~\cite{remark2} which correspond to total 
capacitances of about $C_{\Sigma} \cong
47$~aF. 

Taking into account the electron density, the 
number of electrons in the dots is estimated to be $15 \pm 1$ with
dot radii of about $r_e \cong 54$~nm. This is in
good agreement with the lithographic dimensions seen in
Fig.~1(a). For the energies of the excited states we obtain the following values: $\epsilon^*_{dot1}
\approx 110~\mu$eV and $\epsilon^*_{dot2} \approx 117~\mu$eV. 
Generally, all experiments are carried out at a cryogenic bath temperature of $T_{b} = 50$~mK. 
In temperature dependent measurements, however, the electronic
temperature saturates at a value of $T_{e}~\cong 110-125$~mK.
Furthermore, we determine the total intrinsic width of the resonances to be
$\Gamma~=~\Gamma _{lj}~+~\Gamma _{rj}~\cong~108~\mu$eV
($j = 1,2$ as in Fig.~1(b)).
Summarizing the results so far, the mesoscopic system can be tuned into a
regime $2E_C = U > \epsilon^*_{dots} \sim \Gamma > k_{B}T_{e}$
in which charge transport is dominated by tunneling through single
particle levels.

In the following section we demonstrate that the coupling 
of the two quantum dots can be tuned into different regimes.
For this purpose we connect $gate_3$ and $gate_4$ to each other
and detect the charging diagram of the double dot at a small source drain bias of
$V_{sd} =~-~(\mu_{source}~-~\mu_{drain})/e~= -20~\mu$V.
Having set the voltage for the inner tunneling barrier to be 
$V_{g1} = -317$~mV and $V_{g2} = -349$~mV, 
we find strong coupling of the two quantum dots (see Fig.~2(a)).
In contrast to previous measurements on parallel double quantum dots, 
we detect the whole rhombic pattern in the charging diagram, 
since both dots are equally connected to
the leads~\cite{coupling,parallel}.
For the electrostatic coupling strengths of the two dots we
find $C_{12}/C^{dot1}_{\Sigma} \cong C_{12}/C^{dot2}_{\Sigma} = 0.43 \pm
0.08$, where $C_{12}$ is the interdot capacitance~\cite{eletrostatic}.
As seen in Fig.~2(b) the charging diagram for the
weak coupling regime, i.e. $V_{g1}~=~-537$~mV and
$V_{g2}~=~-594$~mV, shows resonances intersecting,
i.e. the two dots form an AB-ring~\cite{AB}.
Measuring the variation of the amplitude at the crossing points
of Fig.~2(b) by sweeping a perpendicular magnetic field, we detect
AB-oscillations with a periodicity of $\Delta B \approx 16.4$~mT 
(see inset of Fig.~2(b)).
This corresponds to an area of $A = 2.52 \times 10^{-13}$~m$^2$ in 
good accordance with the lithographic size of the two-path dot system.

The fundamental question now arising is to what extent the excess electrons which contribute
dominantly to the current through the whole system are independent of the
core shell structure of the artificial atoms.
For this purpose a different pair of gate voltages is employed
to detect charging diagrams similar to Fig.~2, i.e. $V_{g3}$ and $V_{g4}$.
Hereby, we are able to verify the transition from strong to weak coupling of the
two dots in a more sensitive way.
For an intermediate coupling regime Fig.~3(a) shows such a charging diagram. 
The voltage for the inner tunneling barrier is 
set to be~$V_{g1} = -448$~mV and $V_{g2} = -453$~mV while the
voltage for the two tunneling barriers which define $dot_2$ is tuned to be
$V_{g5} = -854$~mV.
From the charging diagram we can extract the regions with fixed
electron numbers for $dot_1$ and $dot_2$ ($N_1$, $N_2$), which 
is depicted by black lines in Fig.~3(b).
Naturally, the charging diagram is interpreted in terms of capacitances
between the dots and the electrostatic environment,
respectively. In Fig.~3(a) the black line confined by two circles represents
the electrostatic coupling of the two quantum dots.
We obtain electrostatic coupling strengths 
of about $C_{12}/C^{dot1}_{\Sigma} 
\cong  C_{12}/C^{dot2}_{\Sigma} = 0.37 \pm 0.08$.

Apart from the boundaries defined by 
the orthodox electrostatic model, we observe resonances which follow in
parallel to the main resonances (sketched by dotted lines in Fig.~3(b)).
Furthermore, we find resonances which are leaking from a ground state
into the Coulomb blockade regions of the phase
diagram, e.g. the  resonance line between the compartments ($N_1-2,N_2$) and ($N_1-1,N_2$)
can be traced into the ($N_1-1,N_2-1$)-region (dotted-dashed line in Fig.~3(b)). 
For two very large quantum dots similar effects have 
been observed in a sample geometry in which the two quantum dots were  
connected to different drain/source contacts~\cite{parallel2}. 
As was shown in Ref.~\cite{cotunneling} this corresponds 
to higher order tunneling events, indicating strong wave function coupling
of the dots even at low-bias voltages.

In the following we will focus on the tunnel split resonances at the
triple points in the phase diagram marked by the circles
in Fig.~3(b) at $A, B, C, D,$ and $E$~\cite{triplepoints}, 
where the wavefunction coupling is maximum. 
We recorded charging diagrams similar to the one in Fig.~3(a) applying 
a perpendicular magnetic field in the range $B = 0$~T to 2~T~\cite{remark3}. 
Subsequently, conductance traces crossing the split resonances 
are fitted by derivatives of the Fermi-Dirac distribution function with 
respect to $V_{g4}$ (see Fig.~4(a)). 
Fitting the curves in accordance with the maximum tunnel splitting
results in a magnetic field dependence of the splitting which is
depicted in Fig.~4(b)~\cite{figfour}.

Starting with a maximum value at $B = 0$~T all curves follow a characteristic
signature: a minimum around $0.12 - 0.4$~T and a second
maximum at $\sim 0.78 - 1.05$~T. 
For $B > 1.4$~T we find the saturation value of the splittings 
to be $\Delta \epsilon _{s} = 100 - 110~\mu$eV.  
We assume that both an interdot capacitance and an effective overlap 
of the wavefunctions have to be taken into account at the same
time~\cite{QDM}. 
At zero magnetic field both contributions are superimposed. Increasing the
magnetic field the two wavefunctions in the quantum dots are compressed 
and thus their overlap reduced. 
In this model the pure capacitive coupling results in an offset of 
about $\Delta \epsilon _{s} \cong 110~\mu$eV. 
Below $B = 2$~T the curves resemble the magnetic field dependence 
of the Heisenberg exchange energy~$J$ for two excess electrons, 
one in each quantum dot~\cite{Burkhard}. 
Although the main characteristics of all curves in Fig. 4(b) are similar, 
the magnitude of the splitting seems
to be dependent on the specific electron number. Furthermore, the trace which 
corresponds to the triplepoint $B$ lacks a second maximum.
Accordingly, we infer that the coherent coupling of the two quantum dots not only depends on 
the shape of the total wavefunction of two coupled excess electrons as assumed so far,  
but on the specific spin and orbital electron configuration of the whole artificial molecule.
For magnetic fields larger than $B~\geq~B_0~=~0.6$~T 
the Zeeman energy exceeds $\Delta \epsilon_s$ ($\epsilon_{Z} =~(g \mu _B  B)/2$, 
where $g = -0.44$ in GaAs and $\mu _B$ the Bohr magneton). 
Hence, we conclude that the coherent coupling of the 
magnetic field to the electron spins and its effect on the 
wavefunction overlap has also to be taken into account 
in a full theoretical description. 

In summary we have realized an experimental setup 
by which electrons can tunnel through two quantum 
dots in an Aharonov-Bohm geometry, while the coupling $J$ between 
the dots can be broadly tuned. We demonstrate for weakly coupled dots 
that the setup allows to probe Aharonov-Bohm oscillations. 
In an intermediate coupling regime we determine the 
coherent coupling of the two quantum dots 
and extract the magnetic field dependence of the tunnel splitting. 
We conclude that in addition to the coupling 
of the electron spins to an applied magnetic field
the whole shell structure has to be taken into account 
to describe the coherent coupling of the two artificial atoms.

We like to thank H. Qin, J.P. Kotthaus, W. Zwerger,
M. Suhrke, J. K\"onig, and S. Ulloa for helpful discussions and support.
Funds by the Deutsche Forschungsgemeinschaft
(DFG) within the Sonderforschungsbereich SFB 348 
and by the Bundesministerium f\"ur Forschung und Technologie (BMBF)
are gratefully acknowledged.

$^*$~mail: 
{\it Alex.Holleitner@physik.uni-muenchen.de}

\newpage

\begin{figure}[t]
\begin{center}
\caption[fig1]
{
(a)
The device is defined by electron-beam writing in a two step process.
In addition to the conventional Schottky gates defining the quantum dots,
two regions are covered with the resist {\it Calixarene} (see text for details).
(b) By applying appropriate negative voltages to the gates, a
two-path quantum dot system is realized. An electron
in the source contact can tunnel via both dots into the drain contact.
Furthermore, the coupling between the two dots can be tuned by
voltages which are applied to $gate_1$ and $gate_2$.
}
\label{fig1}
\end{center}
\end{figure}

\begin{figure}[t]
\begin{center}
\caption[fig2]
{
The grayscale plots of the charging diagram of the double dot reveal that
the coupling between the two dots can be tuned into different regimes:
(a) As a characteristic of strong electrostatic coupling we find a rhombic
pattern in the charging diagram. Here, the molecular states are spread over the whole double dot.
(white~$\le~I = 0$~pA$~<~$black$~<~I = 16$~pA~$\leq~$white).
(b) In the case of weak coupling we find resonances of the double
dot intersecting (only part of the total charging diagram is shown). 
The device operates as an Aharonov-Bohm (AB) interferometer
(white~$\leq~I~=~0$~pA;~black~$\geq~I~=~9$~pA).
If a magnetic field is applied perpendicular to the quantum dots, the
amplitude of the crossing points produces AB-oscillations as shown in the
inset.
}
\label{fig2}
\end{center}
\end{figure}

\begin{figure}[p]
\begin{center}
\caption[fig3]
{
(a) Charging diagram spanned by $V_{g3}$ and $V_{g4}$
in a coupling regime with strong wavefunction overlap in a logarithmic
gray scale plot (white $\leq I = 0$ pA$ < $black$ < I = 18$ pA$ \leq $white). 
The black line confined by two circles denotes the electrostatic 
coupling of the two quantum dots.
As an indication of the coherent coupling 
the triplepoints~\cite{triplepoints} are split into two resonances 
(exemplarily indicated by two arrows).
(b) The different compartments in
the charging diagram are labeled by possible electron
configurations of the double quantum dot ($N_1$,$N_2$).
Excited states are sketched by dashed and dotted lines, while the
crossing points where the tunnel splitting occurs are marked by
$A, B, C, D,$ and~$E$ (see text for details).
}
\label{fig3}
\end{center}
\end{figure}

\begin{figure}[b]
\begin{center}
\caption[fig4]
{
(a) The logarithmic line plot shows the single trace which is indicated in Fig.~3(a) 
by two black arrows. The tunnel split resonances of point $A$
with respect to $V_{g4}$ can clearly be seen (open boxes) . 
The black lines are fits obtained with derivatives of the Fermi-Dirac distribution -- the
splitting $\delta V_{g4}$ is denoted by a black arrow. 
(b) Magnetic field dependence of $\delta V_{g4}$: Points $A$ to $F$ 
($F$ out of range in Fig.~3) are the crossing points in
Fig.~3(b)~\cite{figfour}. The overall error bar is indicated by
$\pm \delta \epsilon = 13~\mu$eV (see text for details).
}
\label{fig4}
\end{center}
\end{figure}


\begin{thebibliography}{0}
\bibitem{remark1} The pinch-off voltage for the `carbon-bridges` is found
to be 
$V_{g1} = V_{g2} \cong -685$~mV, whereas the presented measurements are
performed at voltages of $V_{g1}
= V_{g2} \le -600$~mV.
\bibitem{remark2} In this section all given values for $dot_1$
were obtained at $V_{g1} = -520$~mV and $V_{g2} = -580$~mV,
whereas  for $dot_2$ $V_{g1} = -530$~mV and $V_{g2} = -600$~mV.
\bibitem{triplepoints} A triplepoint is the point where three charge
configurations of the double quantum dot are degenerate.
\bibitem{figfour} The triple points $A$ to $E$ are defined in Fig.~3(b) --
point $F$ corresponds to the triple point of the
two phase boundaries ($N_1$, $N_2 +1$) / ($N_1 +1$, $N_2+1$)
and ($N_1$, $N_2 +1$) / ($N_1$, $N_2 +2$).
\end{thebibliography}
\end{document}